\documentclass[12pt]{article}
\usepackage{amsmath}
\usepackage{amsfonts}
\usepackage{amssymb}
\usepackage{latexsym}
\usepackage{graphicx}
\usepackage[english]{babel}
\input epsf.sty

\newcommand{\be}{\begin{equation}}
\newcommand{\ee}{\end{equation}}
\newcommand{\ba}{\begin{array}}
\newcommand{\ea}{\end{array}}
\newcommand{\bqa}{\begin{eqnarray}}
\newcommand{\eqa}{\end{eqnarray}}

\begin{document}

\title{
{Oscillating universe with quintom matter}}
 \vspace{3mm}
\author{{Hua-Hui Xiong$^{1}$, Yi-Fu
Cai$^{1}$, Taotao Qiu$^{1}$,}\\ {Yun-Song Piao$^{2}$, Xinmin
Zhang$^{1,3}$}\\
{\small $^{1}$ Institute of High Energy Physics, Chinese Academy of Sciences,}\\ {\small P.O. Box 918-4, Beijing 100049, P. R. China}\\
{\small $^{2}$ College of Physical Sciences, Graduate School of
Chinese Academy of Sciences,}\\ {\small YuQuan Road 19A, Beijing 100049, P. R. China}\\
{\small $^{3}$  Theoretical Physics Center for Science Facilities
(TPCSF), CAS, P. R. China} }
\date{}
\maketitle

\vspace{5mm}

\begin{abstract}

In this paper, we study the possibility of building a model of the
oscillating universe with quintom matter in the framework of
4-dimensional Friedmann-Robertson-Walker background. Taking the
two-scalar-field quintom model as an example, we find in the model
parameter space there are five different types of solutions which
correspond to: (I) a cyclic universe with the minimal and maximal
values of the scale factor remaining the same in every cycle, (II)
an oscillating universe with its minimal and maximal values of the
scale factor increasing cycle by cycle, (III) an oscillating
universe with its scale factor always increasing, (IV) an
oscillating universe with its minimal and maximal values of the
scale factor decreasing cycle by cycle, and (V) an oscillating
universe with its scale factor always decreasing.

\end{abstract}

\newpage

\section{Introduction}

The quintom scenario of dark energy firstly proposed in Ref.
\cite{Feng:2004ad} for the purpose of understanding the dynamical
feature with the equation-of-state (EoS) crossing over the
cosmological constant boundary $w=-1$ differs from the
quintessence or phantom or other scenario of dark energy in the
determination of the evolution of the Universe. In Ref.
\cite{Cai:2007qw}, four of us (Cai, Qiu, Piao and Zhang) have
considered an application of the quintom matter in the early
universe and interestingly we have found a bouncing solution
within the framework of the standard 4-dimensional
Friedmann-Robertson-Walker (FRW) background. In our model the
background evolution can be studied analytically and numerically.
Later on its perturbation theory has been developed in Refs.
\cite{Cai:2007zv,spectrum} and it is found that the perturbations
of this model possess some features of both singular bounce and
non-singular bounce models. In this paper we extend our study to
constructing a model of oscillating universe with the quintom
matter.

The idea of cyclic universe was initially introduced in 1930's by
Richard Tolman \cite{tolman}. Since then there have been various
proposals in the literature. The authors of Refs.
\cite{Khoury:2001wf,Steinhardt:2001vw} introduced a cyclic model
in high dimensional string theory with an infinite and flat
universe. With a modified Friedmann equation the cyclic evolution
of the universe can also be realized
\cite{Brown:2004cs,Baum:2006nz}. In Ref. \cite{Xiong:2007cn} four
of us (Xiong, Qiu, Cai and Zhang) realized that  in the framework
of loop quantum cosmology (LQC) a cyclic universe can be obtained
with the quintom matter. In this paper, however, we will study the
solution of oscillating universe in the absence of the
modification of the standard 4-dimensional Einstein Gravity with a
flat universe.

To begin with, let us examine in detail the conditions required
for an oscillating solution. The basic picture for the evolution
of the cyclic universe can be shown below:
\begin{eqnarray}\label{btb}
... {\rm bounce} \xrightarrow{expanding} {\rm turn}\texttt{-}{\rm
around} \xrightarrow{contracting} {\rm bounce} ...~.
\end{eqnarray}
In the 4-dimensional FRW framework the Einstein equations can be
written down as:\begin{eqnarray}\label{einstein}
 H^2\equiv (\frac{\dot a}{a})^2=\frac{\rho}{3M_p^2}~~~ {\rm and}~~~~
\frac{\ddot a}{a}=-\frac{\rho+3p}{6M_p^2}~,
\end{eqnarray}
where we define $M_p^2\equiv\frac{1}{8\pi G}$, $H$ stands for the
Hubble parameter, while $\rho$ and $p$ represent the energy
density and the pressure of the universe respectively. By
definition, for a pivot (bounce or turn-around) process to occur,
one must require that at the pivot point $\dot a=0$ and $\ddot
a>0$ around the bouncing point, while $\ddot a<0$ around the
turn-around point. According to Eq. (\ref{einstein}), one can get
\begin{equation}\label{condition}
\rho=0~,~~~p<0~~({\rm or}~~p>0)~~~~~{\rm for~the~bounce}~~({\rm
or}~~{\rm turn}\texttt{-}{\rm around}),
\end{equation}
or equivalently, $w\equiv \frac{p}{\rho}\rightarrow -\infty~~({\rm
or}~+\infty)$ at the bounce (or turn-around) point with the
parameter $w$ being the EoS of the matter filled in the universe.
We can see that, when the universe undergoes from bounce to
turn-around, the EoS of the matter evolves from $-\infty$ to
$+\infty$; while in the converse case, the EoS goes from $+\infty$
to $-\infty$. That shows $w$ needs to cross over the cosmological
constant boundary ($w=-1$) in these processes, which interestingly
implies the necessity to have the quintom matter for the
realization of the oscillating universe under the 4-dimensional
Einstein Gravity.

Another interesting evolution of an oscillating universe is that,
this universe undergoes accelerations periodically. In this
scenario, we are able to unify the early inflation and current
acceleration of the universe, leading to the oscillations of the
Hubble constant and a recurring universe. During this kind of
evolution, the universe would not encounter a big crunch nor big
rip. The scale factor keeps increasing from one period to another
and so leads to a highly flat universe naturally. This scenario
was firstly proposed by Ref. \cite{Feng:2004ff} in which a
parameterized Quintom model was used, and in that paper the
coincidence problem was argued to be reconciled.

In this paper we will take the two-scalar-field quintom model for
a detailed study on the oscillating universe. We will show that
for those models considered in Ref. \cite{Cai:2007qw} with a
positive-definite potential there is no oscillating solution, but
it happens when allowing the potential to possess negative
regions. As is proven below, for the two-scalar-field quintom
models, a negative potential is necessary for the scenario of a
cyclic universe. Our paper is organized as follows. In section 2,
we provide an exact solution of the oscillating universe in the
quintom model of two-scalar-field and study the trajectory of the
fields in the phase space. Moreover, we study various possible
evolutions of the universe in our model. Section 3 contains the
conclusion and discussions.

\section{A solution of oscillating universe in the two-scalar-field quintom
model}

\subsection{The model}
The simplest quintom model consists of two scalars with one being
quintessence-like and another the phantom-like. Its action is
given by
\begin{eqnarray}
S=\int d^{4}x \sqrt{-g}\left [ \frac{1}{2}\partial_{\mu}\phi
\partial^{\mu}\phi-\frac{1}{2}\partial_{\mu}\psi\partial^{\mu}\psi-V\left(\phi,\psi\right)
\right ]~,
\end{eqnarray}
where the metric is in form of $\left(+,-,-,-\right)$. In the
framework of FRW cosmology, we can easily obtain the energy density
and the pressure of the model,
\begin{eqnarray}\label{energypressure}
\rho=\frac{1}{2}\dot\phi^2-\frac{1}{2}\dot\psi^2+V(\phi,
\psi)~~,~~~p=\frac{1}{2}\dot\phi^2-\frac{1}{2}\dot\psi^2-V(\phi,
\psi)~,
\end{eqnarray}
and the equations of motion for these two fields,
\begin{eqnarray}
\label{eom1}\ddot\phi+3H\dot\phi+V_{,\phi}=0~~,~~~
\label{eom2}\ddot\psi+3H\dot\psi-V_{,\psi}=0~.
\end{eqnarray}

Phenomenologically, a general form of the potential for a
renormalizable model includes operators with dimension 4 or less
including various power form of the scalar fields. For the study of
this paper we impose a $Z_2$ symmetry on this model, {\it i.e.}, the
potential will keep invariant under the transformations
$\phi\rightarrow-\phi$ and $\psi\rightarrow-\psi$ simultaneously.
Then the potential of the model is given by
\begin{eqnarray}
V(\phi,\psi)&=&V_0+\frac{1}{2}m_1^2\phi^2+\frac{1}{2}
m_2^2\psi^2+\gamma_1\phi^4+\gamma_2\psi^4\nonumber\\
&&+g_1\phi\psi+g_2\phi\psi^3+g_3\phi^2\psi^2+g_4\phi^3\psi.
\end{eqnarray}

From the condition (\ref{condition}), we can see that at both the
bounce and the turn-around point $\dot\psi^2=\dot\phi^2+2V$. From
Eq. (\ref{energypressure}) we have $p=-2V$. When the universe
undergoes a bounce the pressure is required to be negative, which
implies the potential to be positive. However, when a turn-around
solution takes place, the pressure of the universe is required to
be positive, consequently the potential must be negative. The
argument above shows the expected quintom model needs to contain a
negative term in its potential to realize the oscillating
scenario.

In our scenario the two scalar fields dominate the universe
alternately. To show it, let us assume the universe starts an
expansion from a bounce point without losing generality. We find
that, initially the universe is phantom dominated and the energy
density of the universe grows up. However, since we need to enter
a quintessence dominated stage, the energy density has to reach a
maximal value and then decreases. When it arrives at zero, the
turn-around happens and the universe enters into a contracting
phase. Therefore, from the bounce to the turn-around the universe
needs to transit from the phantom dominating phase into the
quintessence dominating phase, or vice versa. In order to describe
the whole evolution explicitly, we in Fig. 1 sketch the evolutions
for the energy density and the scale factor of the cyclic universe
in our model. We can see that during each cycle the universe is
dominated by quintessence $\phi$ and phantom $\psi$ alternately.
However, as is pointed out in Refs.
\cite{Guo:2004fq,Zhang:2005eg}, this process will not happen if
the two fields are decoupled. Therefore, in our model an
interaction between the two fields is also crucial.

\begin{figure}[htp]
\centering
\includegraphics[scale=0.8]{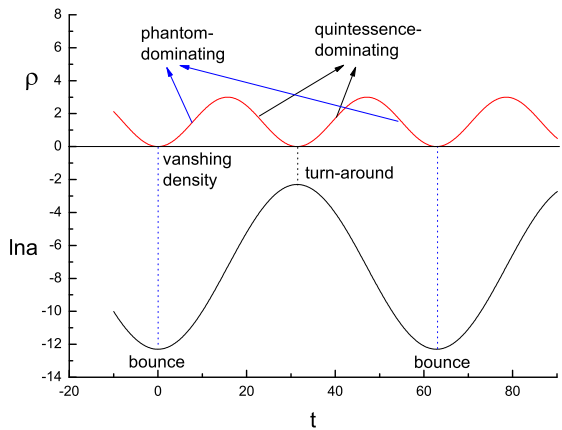}
\caption{ This sketch plot shows the evolution of the energy
density in a cyclic universe. For each cycle the quintessence-like
and phantom-like components dominate alternately. }\label{draft}
\end{figure}

For a detailed quantitative study we take the following form of
the potential
\begin{equation}\label{potential}
V\left(\phi,\psi\right)=\left(\Lambda_0+
\lambda\phi\psi\right)^{2}+\frac{1}{2}m^{2}\phi^{2}-
\frac{1}{2}m^{2}\psi^{2}~,
\end{equation}
where $\lambda$ is a dimensionless constant describing the
interaction between two scalar fields, and $\Lambda_0$ an constant
with dimension of $[{\rm mass}]^2$.

This potential takes a negative value when $\phi$ is near the
origin and  $\psi$ with a large value. However, due to the
interaction between these two fields, the potential is still
bounded from below and is positive definite when the fields are
both away from the center, especially for
$\lambda\phi>\frac{\sqrt{2}}{2}m$.

With the potential (8) we solve the equations of motion
(\ref{eom1}). Inserting the energy density and pressure
(\ref{energypressure}) into the Friedmann equations
(\ref{einstein}), we find a solution where $\phi$ and $\psi$ are
given by:
\begin{equation}\label{ansatz}
\phi=\sqrt{A_{0}}\cos mt~,~~\psi=\sqrt{A_{0}}\sin mt~,
\end{equation}
where the parameter $A_{0}$ describes the oscillating amplitude of
the fields (also with dimension of $[{\rm mass}]^2$). Besides, we
obtain that for this solution $\lambda=\frac{\sqrt{3}m}{2M_p}$, thus
the detailed behavior of the solution is characterized by the three
free parameters which are $m, \Lambda_0, A_{0}$ respectively.

From Eq. (9), we can learn that the two scalar fields oscillate,
however with a phase difference $\pi/2$. So the fields $\phi$ and
$\psi$ dominate the universe alternately, which is what we have
analyzed previously.

\subsection{Classifications of the solutions}

In this section, we study the detailed cosmological evolutions of
this model. We will see that this model gives rise to five
different scenarios of oscillating cosmology within the model
parameter space.

First of all, from the Einstein Equation (\ref{einstein}) one can
easily get the Hubble parameter as follows,
\begin{equation}\label{H}
H=\frac{\sqrt{3}}{3M_p}\left(\Lambda_0+\Lambda_1\sin{2mt}\right)~,
\end{equation}
where we define $\Lambda_1\equiv\frac{\sqrt{3}m}{4M_p}A_0$. For
different parameters, the evolution of the universe can be
classified into five cases. The first one describes an exactly
cyclic universe with its scale factor oscillating periodically.
The second one is that the scale factor of the universe oscillates
with the center value increasing gradually. In the third one, the
Hubble parameter is oscillating periodically but always positive,
correspondingly the scale factor is always increasing during the
whole period of evolution. The fourth one can be viewed as a
reverse process of the second one since in that case the center
value of the scale factor is decreasing gradually. The last case
describes the reverse process of the third one, and in this case,
the scale factor always decreases and so it corresponds to a
nonphysical one. In the following we will study in detail these
five cases one by one. Note that we will take the natural unit
$M_p=1$ in the following calculations.

\begin{figure}[htp]
\centering
\includegraphics[scale=0.6]{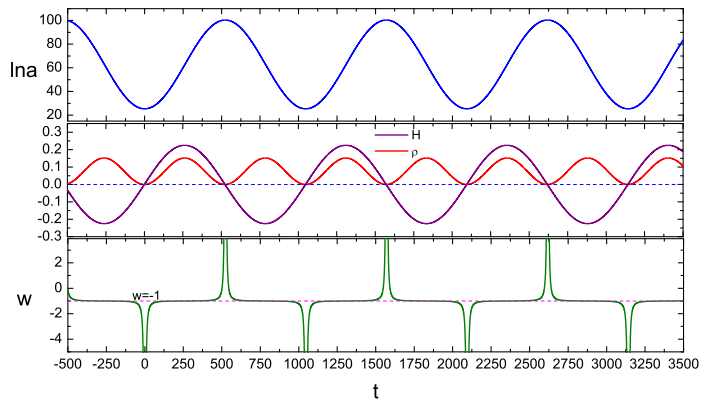}
\caption{ The plot of the evolutions for Case (I). This plot shows
an exactly cyclic universe. The scale factor of the universe
oscillates between the minimal and maximal value. In the numerical
calculation we take $m=3\times 10^{-3}$ and $A_{0}=300$ or
equivalently $\Lambda_1\simeq 0.39$. The Hubble parameter is shown
by the purple line, and the energy density is described by the red
 line.}\label{fig1}
\end{figure}

Case (\uppercase\expandafter{\romannumeral1}): $\Lambda_0=0$. In
this case the Hubble parameter is given by
$H=\frac{mA_{0}}{4M_{p}^{2}}\sin{2mt}$. So for the scale factor we
have
\begin{equation}
\ln{a}\propto \cos{2mt}~.
\end{equation}
This gives rise to a cyclic universe with the minimal and maximal
values of the scale factor remaining the same in every cycle. In
this scenario there is no spacetime singularity.  In Fig.
\ref{fig1} we plot the evolutions of the scale factor, the Hubble
parameter, the energy density and the EoS.

\begin{figure}[htp]
\centering
\includegraphics[scale=0.6]{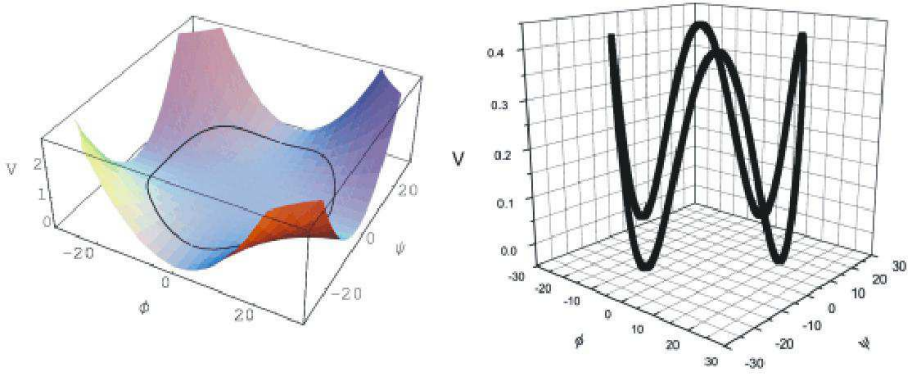}
\caption{The left panel is the potential $V(\phi~,\psi)$ with
respect to the fields $\phi$ and $\psi$, and the right panel is a
closer view of the trajectory of the fields on the potential
surface. The saddle point lies on the zero point in the phase
space. The black solid line shows the trajectory of the fields
$\phi$ and $\psi$ on the surface of the potential.
}\label{potentialfig}
\end{figure}

To show this solution more explicitly, we give the potential $V$
as a function of fields $\phi$ and $\psi$ in Fig.
\ref{potentialfig}. It is easy to see that the potential is
bounded from below in the phase space when $\phi$ is far away from
the origin, and only the center region of the surface is a
hyperbolic paraboloid. In the right panel of Fig.
\ref{potentialfig}, we also give the trajectories of the fields
$\phi$ and $\psi$ on the surface of the potential. From Fig.
\ref{potentialfig} we read that the projection of the trajectory
on the field plane $(\phi, \psi)$ is a circle which is consistent
with the solution (\ref{ansatz}).

\begin{figure}[htp]
\centering
\includegraphics[scale=0.6]{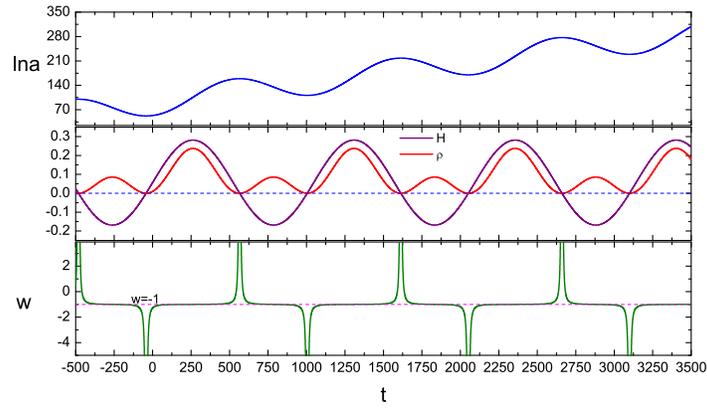}
\caption{ The plot of the evolutions for Case (II). We take the
values of $m$ and $A_0$ to be the same as those used in Fig.
\ref{fig1}, and $\Lambda_0=0.10$ which satisfies the condition
$0<\Lambda_0<\Lambda_1$. This figure also describes an oscillating
evolution of the universe, of which the average value of scale
factor is increasing during the evolution. The figure for energy
density and Hubble parameter is like Fig. \ref{fig1}, however, the
difference is that the oscillating amplitude of the energy density
in the expanding period is larger than in the contracting phase.
}\label{fig2}
\end{figure}

Case (\uppercase\expandafter{\romannumeral2}):
$0<\Lambda_0<\Lambda_1$. In this case the scale factor is solved as
\begin{equation}
\ln{a}\propto C_{1}t+C_{2}\cos{2mt} \label{a}~,
\end{equation}
where $C_{1}$ and $C_{2}$ are constants,
$C_{1}=\Lambda_0/\sqrt{3}M_{p}$, $C_{2}=-A_{0}/8M_{p}^{2}$. The
evolution picture is shown in Fig. {\ref{fig2}}. The solution also
describes a cyclic universe. However, different from Case
(\uppercase\expandafter{\romannumeral1}), as the universe evolves
both the minimal and maximal values of scale factor increase cycle
by cycle. Therefore, the average size of the universe is still
growing up gradually without big crunch or big rip singularities,
although its scale factor experiences contractions and expansions
alternately. With the backward evolution the scale factor can not
shrink to zero in finite time.

\begin{figure}[htp]
\centering
\includegraphics[scale=0.6]{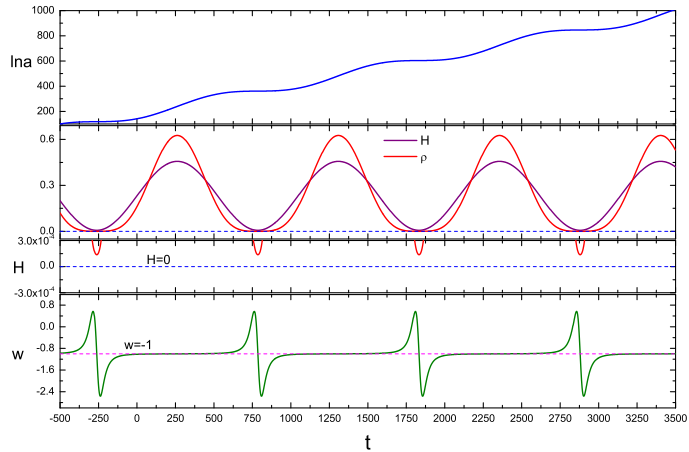}
\caption{ The plot of the evolutions for Case (III). The
coefficients $m$ and $A_0$ are taken the same values as in Fig.
\ref{fig1}, and $\Lambda_0=0.402$ which satisfies the condition
$\Lambda_0\geq\Lambda_1$. This figure shows that the universe is
expanding all the time, without any bounce or turn-around. For
such a universe the energy density and Hubble parameter are always
positive. An interesting feature of this scenario is that the
accelerating expansion of the universe is periodical. This plot
also shows the oscillating EoS around ``$-1$". } \label{fig3}
\end{figure}

Case (\uppercase\expandafter{\romannumeral3}):
$\Lambda_0\geq\Lambda_1$. The solution for the scale factor is the
same as in Case (\uppercase\expandafter{\romannumeral2}). However
the evolution of the universe is different. In this case the
universe lies in the expanding period forever and there is no
contracting phase. An interesting point is that the accelerating
expansion of the universe is periodical. This is shown by the
oscillating EoS in Fig. \ref{fig3}. Since the EoS $w$ is
oscillating around ``$-1$", in the evolution the big rip can be
avoided. For reasonable parameters, we will be able to unify the
inflationary period and the late time acceleration together in
this case. This possibility of oscillating quintom had been
considered in Ref. \cite{Feng:2004ff}, however the concrete field
model is not provided there.

\begin{figure}[htp]
\centering
\includegraphics[scale=0.6]{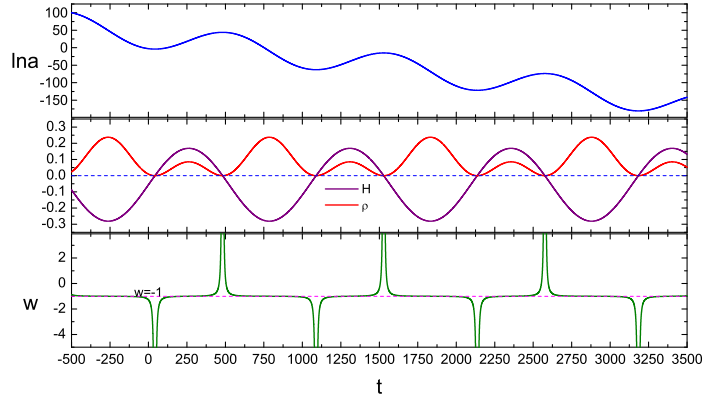}
\caption{ The plot of the evolutions for Case (IV). The values of
$m$ and $A_0$ are also the same as in Fig. \ref{fig1}, and
$\Lambda_0=-0.10$ which satisfies the requirement
$-\big|\Lambda_{1}\big| <\Lambda_0<0$. This figure shows the
oscillating evolution which is contrary to the Fig. \ref{fig2}. In
such an oscillating universe the period of contracting phase is
longer than the expanding phase, so the average value of the scale
factor is decreasing as the universe evolves. \label{fig4} }
\end{figure}

Case (\uppercase\expandafter{\romannumeral4}):
$-\big|\Lambda_{1}\big| <\Lambda_0<0$. This case corresponds to a
cyclic universe with decreasing minimal and maximal scale factor
for each epoch. Contrary to Case
(\uppercase\expandafter{\romannumeral2}), the total tendency is
that the scale factor decreases with the forward cycle. Fig.
\ref{fig4} shows the evolution picture. Moreover, with the forward
evolution the scale factor can not decrease to zero value and
reach the singularity in finite time.

Case (\uppercase\expandafter{\romannumeral5}):
$\Lambda_0<-\big|\Lambda_{1}\big|$. This case describes a
contracting universe for all the time, which corresponds to the
reverse of Case (III). Unfortunately, such a scenario do not give
an expanding evolution and so has already been ruled out by
observations.

Up to now, we have learned that there are mainly five types of
evolutions for our model. Here we would like to naively comment on
the relative merits of these solutions and their physical
significance. First of all, solution (I) shows an exact cyclic
universe and the minimal value of its scale factor does not
vanish, so this solution give the promise of the universe being
singularity-free during the whole evolution. Moreover, for
solutions (II) and (IV), although the minimal value of the scale
factor is approaching to zero in the early time or late time
respectively, this process needs infinite long time. Besides,
since in these two solutions the null energy condition can be
violated by Quintom, there can be no ending point for the geodesic
curve. Therefore, we might say that both the big bang and big
crunch singularities could be avoided in the these three cases.
They can be viewed as one type of oscillating universe in which
the scale factor performs an oscillating behavior. Furthermore,
solution (III) gives another interesting oscillating scenario of
the universe, that is, its Hubble parameter oscillates and keeps
positive. In this scenario, the universe undergoes inflationary
epoch periodically. So it explores an interesting possibility to
unify the early inflation and late time acceleration of the
universe. We do not discuss the last case for it is nonphysical.

\subsection{Constraints on the model parameters}

From the above analysis we know that the possible evolution of the
universe depends upon the value of the constant $\Lambda_0$ in the
potential. For different values of $\Lambda_0$ there exist four
physical possibilities. Three of them contain expansions and
contractions, so describe an cyclic universe. The other one
depicts an forever expanding universe with an EoS oscillating
around the cosmological constant boundary. All of these scenarios
of the cosmological evolutions are free of singularity.

One may notice that the EoS of our model oscillates around the
cosmological constant boundary, and every time when it approaches
$w=-1$ it stays there for a while. Correspondingly, the Hubble
parameter evolves near its maximal value and makes the scale
factor expand exponentially. We treat this period as an
inflationary stage after the bounce. Therefore, in our model there
exists an inflationary period in each cycle. This is an important
process, since the relic matters created in the last cycle could
be washed out during this stage. In this process entropy can be
diluted by inflationary epoch as well, so there is no necessary to
worry about the infinite increase of entropy. Moreover, some of
the primordial perturbations are able to exit the horizon, and
reenter the horizon when the inflationary stage ceases. When these
perturbations reenters, new structures in the next period will be
formed.

Now we study the possible constraints on the model parameters ($m,
\Lambda_0, A_{0}$).  First of all, we require the period of the
oscillation of the Hubble parameter be no less than 2 times the
age of our universe which is of order of the present Hubble time.
So from Eq. (\ref{H}) we can get the period $T=\frac{\pi}{m}\sim
{\cal O}({H_0}^{-1})$ where ${H_0}^{-1}$ is the Hubble time and
${H_0}^{-1}\simeq 10^{60} M_p^{-1}$, so $m\sim {\cal
O}(10^{-60})M_p$. Secondly we require the maximal value of the
Hubble parameter be able to reach the inflationary energy scale.
The maximal value of $H$ is given by ($\Lambda_0+\Lambda_1$) when
($\sin 2mt$) reaches its maximum. For case (I), (III) and (IV),
$|\Lambda_0|\leq\Lambda_1$, so we have $(\Lambda_0+\Lambda_1)\sim
{\cal O}(\frac{m A_0}{{M_p}^2})$. If we consider inflation with
energy density, for example around ${\cal O}(10^{-20}){M_p}^4$, we
find $A_0$ must be of ${\cal O}(10^{50}){M_p}^2$. Such a large
value of $A_0$ indicates the scalar fields $\phi$ and $\psi$ take
values much beyond the Planck scale, which makes our effective
lagrangian description invalid. One possibility of solving this
problem is to consider a model with a large number of quintom
fields. In this case the Hubble parameter is amplified by a
pre-factor $\sqrt{N}$ with $N$ the number of quintom fields. If
$N$ is larger than or the same order of $10^{50}$, $A_0$ can be
relaxed to be ${\cal O}(1)M_p^2$ or less.

\section{Conclusion and discussions}

In this paper, we have studied the possibility of constructing an
oscillating solution of the universe in the two-scalar-field
quintom model. Our results show that this model gives rise to
cyclic cosmology where a universe expands and contracts
alternately without singularity. Within the model parameter space
we also provide a new kind of evolution for the oscillating
universe, i.e, expanding forever with its EoS oscillating around
$-1$ which leads to the universe accelerating periodically.

Cyclic universe scenario has been widely studied in the
literature. Among them, there are cyclic models in the braneworld
scenario \cite{Steinhardt:2001vw,Baum:2006nz}, closed oscillating
universe \cite{Clifton:2007tn}, cyclic universe in Loop Quantum
Cosmology \cite{loop-branecyc,singh-bounce,Xiong:2007cn}, and see
Refs.
\cite{piao:2003cy,oscillating-universe,Baum:2006ee,Freese:2008pu,Zhang:2007yu,Kanekar:2001qd,Dabrowski:2004hx,Biswas:2008ti,Aref'eva:2008gj}
for recent developments on kinds of oscillating universes. The
main difference of our work presented in this paper from theirs is
that the model we considered is restricted within the framework of
4-dimensional Einstein Gravity in a flat universe.

\section*{Acknowledgments}

We are grateful to Robert Brandenberger, Mingzhe Li, Yongchao Lv,
Anupam Mazumdar, and Junqing Xia for helpful discussions. This
work is supported in part by National Natural Science Foundation
of China under Grant Nos. 90303004, 10533010 and 10675136 and by
the Chinese Academy of Science under Grant No. KJCX3-SYW-N2.

\end{document}